\documentclass[apj]{emulateapj}
\usepackage{txfonts}

\bibpunct{(}{)}{;}{a}{}{,}

\newcommand{\Msunyr}{\mbox{$\rm M_{\odot}$\,yr$^{-1}$}}

\newcommand{\Mdot}{\mbox{$\dot{M}$}}

\def\mathstacksym#1#2#3#4#5{\def#1{\mathrel{\hbox to 0pt{\lower#5\hbox{#3}\hss} \raise #4\hbox{#2}}}}
\mathstacksym\gta{$>$}{$\sim$}{1.5pt}{3.5pt} 
\mathstacksym\lta{$<$}{$\sim$}{1.5pt}{3.5pt} 
\begin{document}


\title{A red supergiant nebula at 25 micron: arcsecond scale mass-loss asymmetries of $\mu$ Cep\altaffilmark{1}}

\altaffiltext{1}{Based on data collected at the Subaru telescope, which is operated by the National Astronomical Observatory of Japan.}

\shorttitle{Subaru imaging of $\mu$ Cep}
\shortauthors{de Wit et al.}
\author{W.J. de Wit\altaffilmark{2}, R.D. Oudmaijer\altaffilmark{2}, T. Fujiyoshi\altaffilmark{3}, M.G. Hoare\altaffilmark{2}, M. Honda\altaffilmark{4},
H. Kataza\altaffilmark{5}, T. Miyata\altaffilmark{6}, Y. K. Okamoto\altaffilmark{7}, T. Onaka\altaffilmark{8}, S. Sako\altaffilmark{6}, T. Yamashita\altaffilmark{3}} 
                                %
%
%
%
                                %
\altaffiltext{2}{School of Physics \& Astronomy, University of Leeds, Woodhouse Lane, Leeds LS2 9JT, UK; w.j.m.dewit@leeds.ac.uk}
\altaffiltext{3}{Subaru Telescope, National Astronomical Observatory of Japan, National Institutes of Natural Sciences, 650 North A'ohoku Place, Hilo, HI 96720, USA}
\altaffiltext{4}{Department of Information Science, Kanagawa University, 2946 Tsuchiya, Hiratsuka, Kanagawa, 259-1293, Japan}
\altaffiltext{5}{Department of Infrared Astrophysics, Institute of Space and Astronautical Science, Japan Aerospace Exploration Agency, Sagamihara, Kanagawa, 229-8510, Japan}
\altaffiltext{6}{Institute of Astronomy, University of Tokyo, Osawa 2-21-1, Mitaka, Tokyo 181-0015, Japan}
\altaffiltext{7}{Faculty of Science, Ibaraki University, 2-1-1 Bunkyo, Mito, Ibaraki, 310-8512, Japan}
\altaffiltext{8}{Department of Astronomy, Graduate School of Science, University of Tokyo, Bunkyo-ku, Tokyo 113-0022, Japan}
%
                                %
\begin{abstract}
We present diffraction limited (0.6\arcsec) 24.5\,\micron~ Subaru/COMICS
images of the red supergiant $\mu$ Cep. We report the detection of a
circumstellar nebula, that was not detected at shorter wavelengths. It
extends to a radius of at least 6\arcsec~in the thermal infrared. On
these angular scales, the nebula is roughly spherical, in contrast, it
displays a pronounced asymmetric morphology closer in. We
simultaneously model the azimuthally averaged intensity profile of the
nebula and the observed spectral energy distribution (SED) using
spherical dust radiative transfer models. The models indicate a
constant mass-loss process over the past 1000 years, for mass-loss
rates a few times $10^{-7}$\,\Msunyr.  This work supports the idea that
at least part of the asymmetries in shells of evolved massive stars
and supernovae may be due to the mass-loss process in the red
supergiant phase.
\end{abstract}


\section{Introduction}
\label{intro}
Although the final stages of the post-main sequence evolution of
massive stars do not last long, it is here where most of the mass is
lost and the shaping of the pre-supernova ejecta takes place.  Key in
this respect are the Red Supergiants (RSGs) which represent a phase in
the life of stars with initial masses 10--30\,M$_{\odot}$. During this
phase lasting 10$^4 -10^5$ years, the stars lose prodigious amounts of
mass at a rate of order 10$^{-6} - 10^{-4}$ M$_{\odot}$ yr$^{-1}$ (van Loon et
al. 2005\nocite{2005A&A...438..273V}; Massey et
al. 2008\nocite{2008arXiv0801.1806M}), and the final mass of the
objects is mostly set during this phase. RSGs have been observed to be
the direct progenitors of Type IIP supernovae (Smartt et
al. 2004\nocite{2004Sci...303..499S}), but can also evolve towards the
blue via a Yellow Hypergiant phase (Oudmaijer et
al. 2008\nocite{2008arXiv0801.2315O}; de Wit et
al. 2008a\nocite{2008A&A...480..149D}) to become a Wolf-Rayet star and
eventually explode as a supernova (Meynet \& Maeder
2000\nocite{2000A&A...361..101M}).

It is not only the study of the stars themselves that helps us
understand the final stages of the evolution of massive stars. 
Investigating RSGs' mass-loss and their circumstellar
material helps us to understand the origin of the aspherical
structures found around supernovae, such as SN 1987A's rings, or 
gamma-ray bursts.  It is possible that at least some of
these observed asymmetries originate during the RSG phase (e.g. Chita et al. 2008). Indeed, in
several instances, the mass-loss during the RSG phase has been found
to deviate from spherically symmetric, and can have a complex
appearance. Much effort has been directed towards the objects VY CMa
and NML Cyg, extremely bright, cool, objects close the empirical
Humphreys Davidson limit (Humphreys \& Davidson 1979\nocite{1979ApJ...232..409H}). 
High resolution studies have
revealed highly anisotropic structures in the wind of VY CMa
(e.g. Smith et al. 2001\nocite{2001AJ....121.1111S}), while NML Cyg's
aspherical appearance is shaped by the strong radiation from the
nearby Cyg OB2 association (Morris and Jura 1983; Schuster et al
2006\nocite{2006AJ....131..603S}).

A third RSG that is found in the same location in the HR diagram as
the above two objects is $\mu$ Cep (Schuster et al 2006\nocite{2006AJ....131..603S}). 
Contrary
to these two cooler objects, little is known about the material
surrounding $\mu$ Cep. This can be readily explained by its mass-loss,
which is orders of magnitude lower than for NML Cyg or VY CMa
(cf. Jura \& Kleinman 1990\nocite{1990ApJS...73..769J}).  
With an $M_{\rm bol}$ of $-$9.08, this M2Ia star is one of the brightest RSGs known, lying on
evolutionary tracks corresponding to stars with initial masses greater
than 25 M$_{\odot}$ (Levesque et al. 2005\nocite{2005ApJ...628..973L}).  Using Levesque et al.'s
parameters, we find a distance to the object of 870 pc, and a
luminosity of $\sim$ 3$\times 10^5$ L$_{\odot}$.
Here, we report on the discovery of an extended dust shell around the object by means
of diffraction limited imaging (0.6\arcsec) of mid-IR thermal dust emission. 

\section{Observations and data reduction}
\label{obser}
\begin{figure}
  \epsscale{1.0}
  \plotone{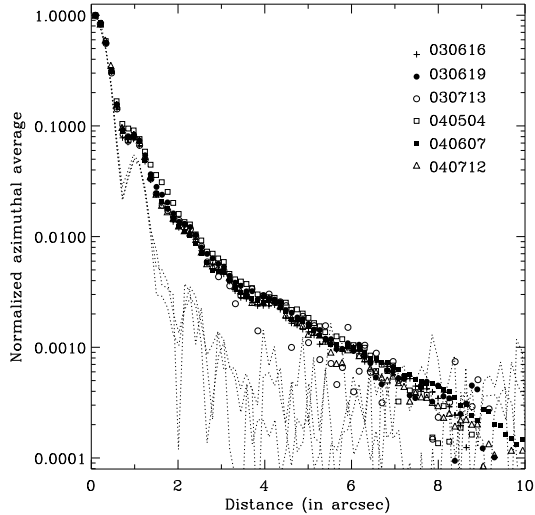}
  \caption[]{The azimuthally averaged and normalized intensity profile of $\mu$ Cep as
  observed at 6 different dates. The profile is stable. Note the slightly narrower profile for the 030713 data. These are  
  less sensitive, and with a SNR a factor of 10 less
  than the other data, the faint outer parts can not be measured. Also shown are the intensity profiles of the 3 genuine PSF objects. Note the first
two Airy rings at $\sim1.2\arcsec$ and $\sim2.4\arcsec$.} 
  \label{obspsf} 
\end{figure}

$\mu$ Cep was observed on 6 different occasions as a mid-IR standard
star between June 2003 and July 2004 with the COMICS instrument
mounted on the 8.2 meter Subaru telescope in Hawaii (Kataza et
al. 2000\nocite{2000SPIE.4008.1144K}; Okamoto et
al. 2003\nocite{2003SPIE.4841..169O}; Sako et
al. 2003\nocite{2003SPIE.4841.1211S}). The star was imaged using the
Q24.5-OLD and Q24.5-NEW filters both centred on 24.5\,\micron. The imaging mode of
COMICS utilises a Raytheon 320$\times$240 Si:As IBC array with a pixel
scale of 0.13\arcsec. This scale comfortably samples the diffraction
limit for the 8.2m Subaru telescope at this wavelength, which is
0.6\arcsec. The observations are summarised
in table\,1.  More details on the filters and data reduction can be
found in de Wit et al. (2008b, in prep.).

$\mu$ Cep is bright and therefore often used as a standard star,
either for determining the instrumental point-spread-function (PSF) or
for calibrating photometry. We observed the object for calibration
purposes for a different project (as reported in de Wit et al. 2008b),
and found that it was extended compared to other PSF reference
stars. As $\mu$ Cep was the PSF standard, we do not have a concomitant
PSF reference observations. However, we can use as a reference PSF
the remaining three genuine PSF objects observed for the same
project: $\alpha$ Tau and asteroids 51 and 511. To illustrate the
stability of the observational set-up and the extent of $\mu$ Cep, we
show in Fig.~\ref{obspsf} the, azimuthally averaged, normalized
intensity profiles of the 3 genuine PSF standards and those of each of
the 6 observations of $\mu$ Cep.  The 3 point sources have virtually
identical profiles, and the noise begins to dominate only after $\sim$4\arcsec.
On the contrary, the profiles for $\mu$ Cep are clearly extended with respect
to the 3 PSF standards to at least 6\arcsec, and they are also
hardly different from each other on all occasions. The lowest signal-to-noise ratio (SNR) data
of $\mu$ Cep (taken in July 2003) also reveal that the object is
extended, but the data are not deep enough to probe the extent as well
as the higher SNR data.  We conclude that $\mu$ Cep is extended at
24.5\,$\mu$m.

\section{Results}
\begin{figure*}
\plottwo{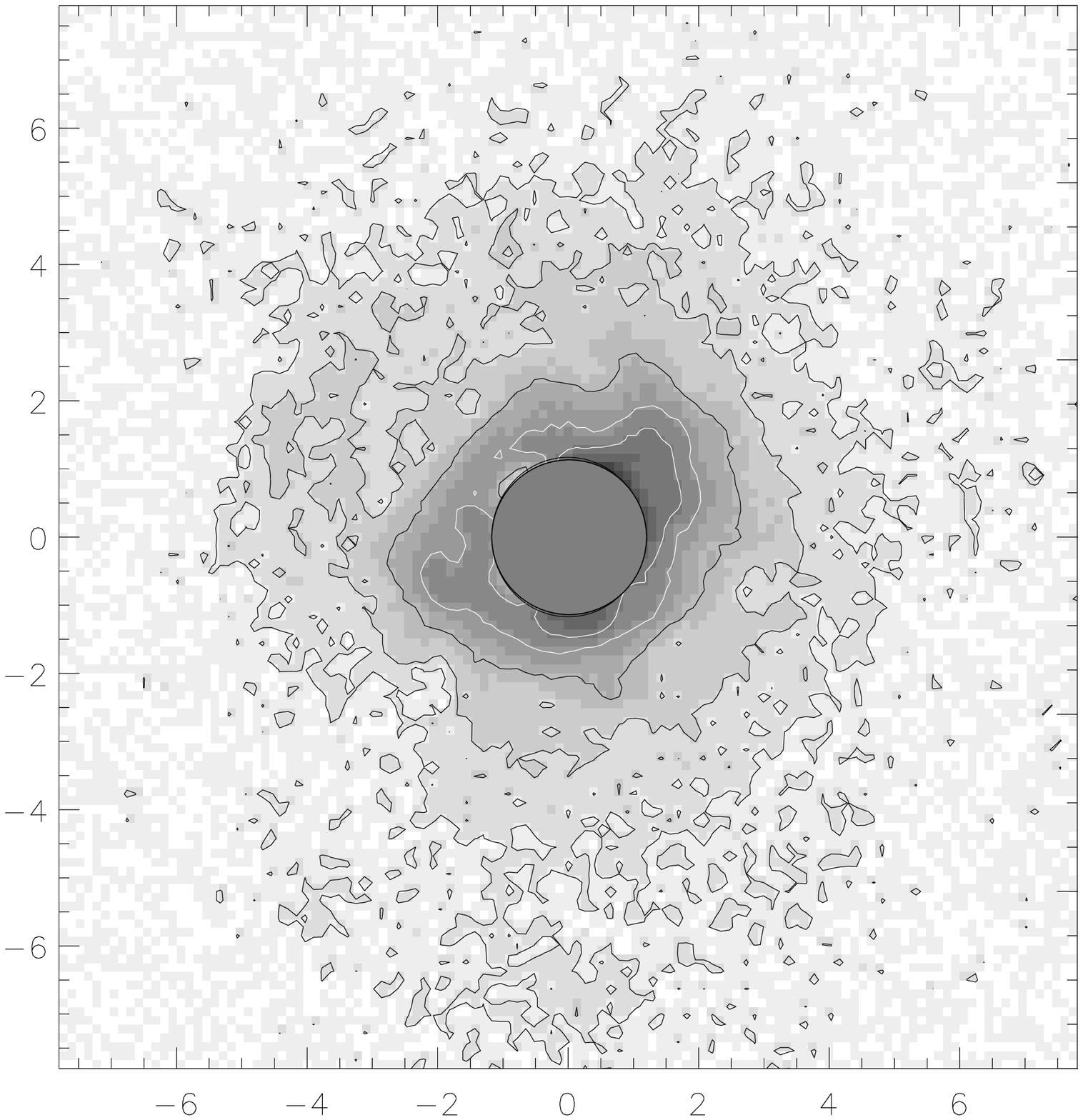}{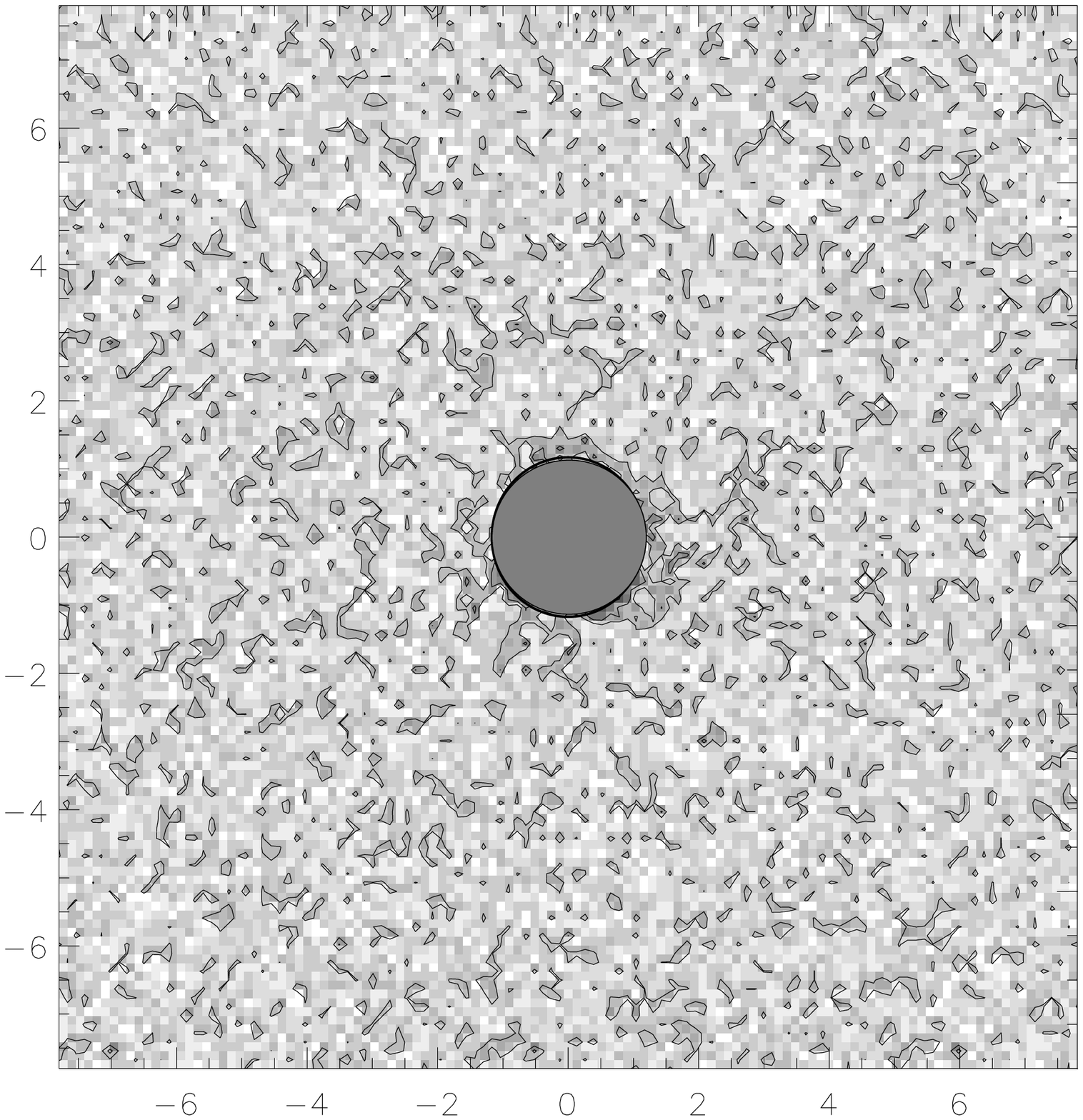}
\caption[]{The PSF subtracted image of $\mu$ Cep (left) and the PSF standard
  $\alpha$ Tau (right). The instrumental PSF was determined from the point
  source asteroid\,511. The region within 1.2\arcsec~is masked to hide
  artefacts from the subtraction procedure. $\mu$ Cep is clearly extended, and
  an asymmetry originates from close to the star. The contour levels are
  chosen to enhance the contrast between the intensity levels, lowest contour corresponds to twice 
  the standard deviation in the background level. North is to the left, East is to the bottom.}
\label{images}
\end{figure*}

\subsection{The 24.5\,$\mu$m images: asymmetric dust distribution}

The highest SNR image (taken on 04/06/07) of $\mu$ Cep is shown in the
left panel of Fig.~\ref{images}. We have subtracted from this image
the scaled profile of the PSF standard asteroid 511 in order to
enhance the features of $\mu$ Cep's circumstellar emission. For
comparison we show in the right panel of Fig.~\ref{images} the PSF standard 
$\alpha$ Tau with again a PSF profile subtracted. It is clear that $\mu$ Cep has extended circumstellar
emission reaching distances from the star of at least 6\arcsec. The
outer parts of the shell-like structure are roughly circular
with indications of larger scale inhomogeneities. In contrast, the inner
parts reveal a clearly asymmetric, bipolar type, geometry at a
position angle of $\sim$67$^{\rm o}$.  The elliptical structure is
visible from the inner parts and extends to 2.2\arcsec~along the major
axis on either side of the central star, and 1.4\arcsec~along the
minor axis.  The intensity contrast between the brighter regions and
the fainter regions at the same distance from the star is a factor of
2.

\subsection{Dust radiative transfer modelling}
\label{sec:mod}
Emission at 24.5\,\micron~stays optically thin for large columns of dust. 
Resolved images allow us to simultaneously model the dust emission as a function
of distance from the star (the intensity profile) and the total dust emission
as given by the SED. For this purpose, we employ DUSTY, a code that
solves self-consistently the scaled 1D dust radiative transfer problem (see
Ivezi\'c \& Elitzur 1997\nocite{1997MNRAS.287..799I}). We use a spherically symmetric dust distribution,
that is illuminated by a central, unresolved star. Its radiation is
represented by a Kurucz ATLAS9 atmosphere model (Kurucz 1993\nocite{1993yCat.6039....0K}) with
$T_{\rm eff}=3750\,{\rm K}$, solar metallicity and surface gravity
corresponding to the supergiant nature of $\mu$ Cep, i.e. $\rm log(g)=0.0$. 
We use oxygen-rich dust with a condensation temperature of 1000\,{\rm K}. The dust particles
follow a MRN size distribution (Mathis, Rumple, \& Nordsieck
1977\nocite{1977ApJ...217..425M}). Within the limitation of a spherical model,
we experiment with (1) the amount of dust as parametrized by the total optical
depth $A_{V}$; (2) the radial density distribution of the
dust $\propto r^{-p}$, adopting a constant ($p=2$) or declining ($p=1.5$) mass-loss rate; (3) two different silicate opacity tables, the ones provided by
Draine \& Lee (1984\nocite{1984ApJ...285...89D}) and for ``warm'' silicates by 
Ossenkopf et al. (1992\nocite{1992A&A...261..567O}). The
outer bound of the model is set at 1000 times the dust condensation radius,
but the exact value is of little influence on the total excess flux at
wavelengths shortward of $\sim$25\,\micron~as long as the outer radius is much larger
than the inner dust condensation radius.

We build $\mu$ Cep's SED using a mid-IR spectrum taken with the short
wavelength spectrometer (SWS, de Graauw et
al. 1996\nocite{1996A&A...315L..49D}) on board the ISO satellite
(Kessler et al. 1996\nocite{1996A&A...315L..27K}), near-IR $JHKLM$
photometry from Heske (1990)\nocite{1990A&A...229..494H} and visual
broadband $UBVRI$ Johnson photometry from Lee (1970\nocite{1970PASP...82..765L}).  The photometry is dereddened
for an interstellar extinction of $A_{V}=1.5^{m}$ 
(Levesque et al. 2005). The IRAS 60\,$\mu$m and 100\,$\mu$m data
points lie above the extrapolated ISO-SWS spectrum. This excess flux
is due to the much larger beam of IRAS and comes from extended cool
dust emission. The origin of this emission is not obvious, it could be
either due to the interstellar medium being heated up by the
stellar radiation (cf. Oudmaijer 1996), or a very extended shell which
is the result of a previous mass-loss phase (e.g. Stencel et
al. 1988). It should be kept in mind that we only consider the most
recent mass-loss episode here. 

The fit procedure initially estimates the $L_{\rm
bol}$ by matching the overall shape of the scaled model SED to the
observed one. DUSTY's output images are then accordingly scaled and
convolved with the instrumental PSF. A comparison of model infrared
excess and model intensity profile to the observed ones is made for
all generated models. A simple tally is performed based on a goodness-of-fit 
criterion. The model that fits both sets of data best is the
one with the highest average ranking in the two tallies.

\begin{figure*}
  \center{\includegraphics[height=8cm,width=6cm,angle=90]{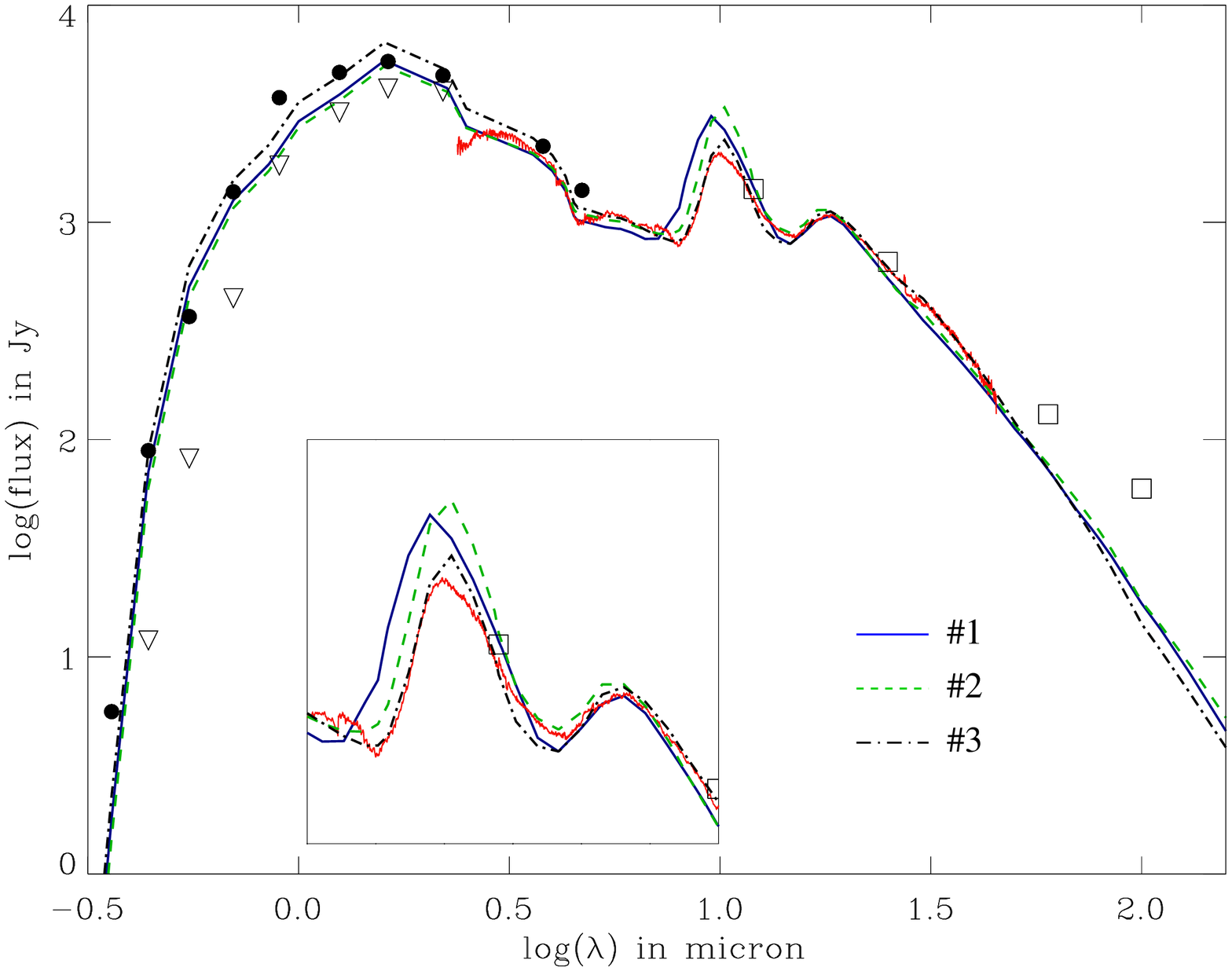}
    \includegraphics[height=8cm,width=6cm,angle=90]{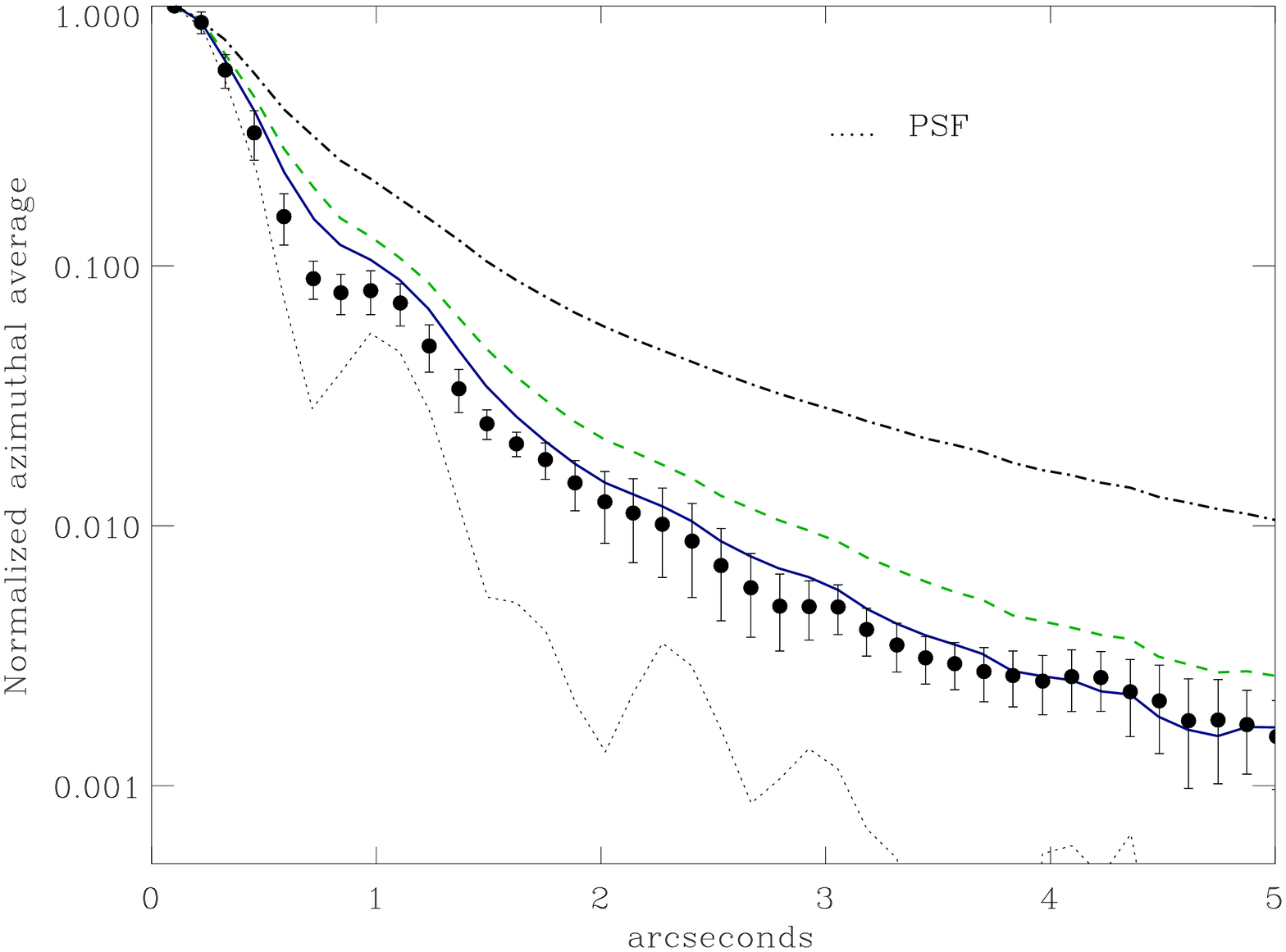}}
  \caption[]{Model fits to the SED (left panel) and the 24.5\,\micron~intensity profile (right panel). Three models are
    shown corresponding to the parameters given in Table\,2. Filled circles represent the extinction corrected optical and near-IR 
data (open triangles). Sources of the SED data are discussed in the text. }
  \label{SED}
\end{figure*}

The results of the 1D modelling are presented in Fig.\,\ref{SED}.
Once the intensity profile is fit for a given
nebular structure and stellar luminosity, the SED can be matched by
increasing the amount of dust in the given nebula. The normalized
intensity profile is quite insensitive to the total amount of dust for
small optical depths. We chose to prioritize a fit to the slope of the
continuum longward of the silicate feature rather than the silicate
emission profile. Details for each of the three models presented
in Fig.\,\ref{SED} are given in Table\,1.  We present the two
best-fitting $p=2$ models corresponding to the two different silicate
opacities. The silicate emission feature is not well matched by
either model, although the intensity profile is better fit by model
\#1 (full line). At the inner 2\arcsec~the models predict too much intensity,
which is consistent with a deviation from spherical symmetry seen in
the images.  Models with a shallow $p=1.5$ radial density profile fit
the intensity profile worst as is illustrated by model \#3. This
demonstrates the value of spatial information in this type of analysis,
because the SED is at the same time very well matched by model
\#3. All models have a mass-loss rate of a few $10^{-7}\,\Msunyr$ for an
expansion velocity of $\rm 10\,km\,s^{-1}$ (cf. Le Borgne \& Mauron
1989\nocite{1989A&A...210..198L}).

\section{Discussion}
We have presented the first high-resolution (0.6\arcsec) images of the
circumstellar environment of the RSG $\mu$ Cep. This material has not been seen
in previous imaging campaigns. For a sample of massive evolved objects,
Schuster et al. (2006) obtained deep optical images with the HST to
search for scattered light by circumstellar dust. They detected only those
sources which have mass-loss rates many orders of magnitude larger than $\mu$ Cep over the past 500--1000
years. Le Borgne \& Mauron (1989) searched for optical, reflected
emission in their polarization data. These authors made pointed
observations with a 10\arcsec~diaphragm at several locations around
the star and detected (faint) polarized light as far as 20\arcsec~from
the star. Other evidence for extended emission comes from Mauron (1997\nocite{1997A&A...326..300M}), who finds
resonance scattered K{ \sc i} line emission at distances further than
20\arcsec~from the star. These data were taken at 2 slit positions,
and thus do not allow for conclusive statements about the geometry of
the circumstellar matter to be made. They do appear to indicate the
presence of either clumpy material or several discrete mass-loss
episodes in the recent past.  The data presented in this paper
traces material out to 6\arcsec. As an expansion velocity of
10\,km\,s$^{-1}$ at a distance of 1 kpc corresponds to about 1\arcsec~per
500 yr, the mass lost 2000--3000 years ago appears to have been ejected 
in a roughly spherically symmetric way. However, the most recent mass lost, ejected less than
1000 years ago, shows a pronounced axisymmetric geometry in our high resolution
images.

\begin{table}
   {
     \begin{center}
       \caption[]{Input parameters and derived quantities of the models presented in Fig\,\ref{SED}. DL and OW dust types stand
for Draine \& Lee (1984), and Ossenkopf et al. (1992) silicates. Model \#1 is the preferred model.}
       \begin{tabular}{ccccclccl}
         \hline
         \hline
         \#    &  p  & $R_{\rm in}$ & $R_{\rm out}/R_{\rm in}$      & $L$         &   dust & $A_{V}$ &   $M_{\rm dust}$ & $\Mdot$  \\
                  &     & (mas)     &    & ($L_{\odot}$)&   &        &  ($ M_{\odot}$)& (\Msunyr)\\
 	\hline            
        1        &  2.0   & 75   & $1\,10^{3}$  & $3.3\,10^{5}$ & DL  & 0.70  &  $1.4\,10^{-2}$  &  $4.5\,10^{-7}$\\
        2        &  2.0   & 103  & $1\,10^{3}$  & $3.1\,10^{5}$ & OW  & 0.60  &  $2.2\,10^{-2}$  &  $5.2\,10^{-7}$\\    
        3        &  1.5   & 112  & $1\,10^{2}$  & $3.9\,10^{5}$ & OW  & 0.30  &  $5.0\,10^{-3}$  &  $1.6\,10^{-7}$\\
 	\hline
       \end{tabular}
     \end{center}
   }
   \label{tab:parameters}
\end{table}

One may speculate whether this geometry is due to a bipolar flow colliding with the
spherically symmetric previous wind, or a slowly expanding torus.
It is useful in this respect to consider the current findings for the
more numerous lower-mass counterparts of RSGs, the
Asymptotic Giant Branch (AGB) stars. Data on these are also sparse,
but asymmetries of the circumstellar material have been seen at high-resolution 
in several cases (e.g. Menut et al. 2007\nocite{2007MNRAS.376L...6M} and Murakawa et
al. 2005\nocite{2005A&A...436..601M} for IRC +10216; 
Vinkovi\'c et al. 2004\nocite{2004MNRAS.352..852V} and Inomata et
al. 2007\nocite{2007PASJ...59..799I} 
for IRC +10011). The picture that emerges from such individual studies
is that the outer parts of the AGB stars are fairly spherical. The inner parts are 
relatively complex, and can be best  explained with a torus type structure 
close to the star, perhaps carved out by a bi-polar flow. 
Based on a compilation of the best data available, Huggins (2007\nocite{2007ApJ...663..342H})
finds that, in general, tori correspond to low outflow velocities and
jets with higher speeds, in excess of 100\,km\,s$^{-1}$. 
In addition, he concludes that the jets
develop shortly after a torus is ejected. Given that the CO spectra of
$\mu$ Cep indicate low velocities rather than high velocities, it is
very well possible that we are now witnessing the same for a Red
Supergiant.

\section{Concluding Remarks}

We have obtained the first diffraction-limited images of a Red
Supergiant at 24.5\,$\mu$m, and resolved the circumstellar material
around $\mu$ Cep out to 6\arcsec. The intensity profile and the SED 
were simultaneously fitted with a dust model. The main results can be summarized as follows:


1. The outer parts of the shell are to first order circular with apparent inhomogeneities, the
  inner parts, tracing the mass lost in the past 1000 years display a
  flattened structure, which is possibly a slowly expanding, dense
  torus. The immediate conclusion from this is that any asymmetries in
  the shells of massive evolved stars and Supernovae ejecta may find their
  origins in the Red Supergiant phase. This is qualitatively very
  similar to what is found for the lower-mass AGB stars which evolve
into bipolar Planetary Nebulae.

2. The mass-loss rate for $\mu$ Cep is found to be a few times 10$^{-7}$\,M$_{\odot}\,$yr$^{-1}$. This low mass-loss rate may explain the fact
  that $\mu$ Cep is not detected in scattered light, as opposed to other
  RSGs which have mass-loss rates that are orders of magnitude larger.
  
3. A general conclusion is that imaging thermal dust emission at
  24.5\,$\mu$m is a viable manner to spatially resolve the
  circumstellar material around evolved objects with low
  mass-loss rates and as a consequence are otherwise not detected with
  maser observations, CO imaging or scattered light imaging.  


\acknowledgements
RDO is grateful for the support from the Leverhulme Trust for awarding
a Research Fellowship. We thank Martin Groenewegen for fruitful discussions.
The version of the ISO data presented in this paper correspond to the 
Highly Processed Data Product (HPDP) set called hpdp\_39802402\_5 
by W.F. Frieswijk et al., available for public use in the ISO Data Archive.

\bibliographystyle{aa}

\end{document}